# Thermal evolution of Pt-rich FePt/Fe$_3$O$_4$ heterodimers studied using x-ray absorption near edge spectroscopy

S. J. A. Figueroa[1], S. J. Stewart[2], T. Rueda[3], A. Hernando[3,4], P. de la Presa[3,4*]

1. European Synchrotron Radiation Facility, 6 rue Jules Horowitz, BP 220, 38043, Grenoble Cedex, France

2. IFLP, CCT-La Plata, CONICET, Departamento de Física, Facultad de Cs. Exactas, Universidad Nacional de La Plata, CC 67, 1900 La Plata, Argentina

3. Instituto de Magnetismo Aplicado, UCM-ADIF-CSIC, P. O. Box 155, 28260 Las Rozas, Spain

4. Departamento de Física de Materiales, Facultad de CC. Físicas, UCM, 28040 Madrid, Spain

*Title running head*:  Thermal evolution of Pt-rich FePt/Fe$_3$O$_4$ heterodimers

\*corresponding author: pm.presa@pdi.ucm.es

**Abstract**: FePt/Fe$_3$O$_4$ nanoparticles can be used as building blocks to obtain, upon thermal annealing, magnetic nanocomposites with combined magnetic properties. Although the pre- and post-annealed samples are usually well characterized, a detailed investigation during annealing is necessary to reveal the role of intermediate processes to produce a desirable composite. We present an alternative method using *in situ* XANES to investigate the thermal evolution of oleic acid and oleylamine coated Pt-rich FePt/Fe$_3$O$_4$ heterodimers. As the temperature increases, a progressive reduction of Fe$_3$O$_4$ to FeO occurs helped by the thermolysis of the surfactants; while above 550 K Fe$_3$Pt starts to be formed. At 840 K an abrupt increase of FeO further drives the phase transformation to stabilize the iron platinum soft phase. Thus, the Fe$_3$O$_4$ reduction acts as catalyst that promotes the Fe and Pt interdiffusion between the Pt-rich FePt and Fe$_3$O$_4$/FeO to form Fe$_3$Pt instead of exchange coupled FePt/Fe$_3$O$_4$ with hard magnetic properties. In addition, the role of the interface of the heterodimer ends is discussed. The pre- and post-annealed samples were also characterized by TEM, XRD, EXAFS, magnetometry and Mössbauer spectroscopy.

**Introduction**

Multicomponent nanostructured materials offer a series of functionalities due to the affording possibilities to retain the physical-chemical properties of single-component counterparts, thus making them attractive for specific purposes.[1] Besides, the coupling interactions between domains of coexisting materials are of relevant importance to define the multicomponent features.[2] These multifunctional nanomaterials can exhibit novel physical and chemical properties, which are essential for future technological applications if the material structure and interface coupling interactions are adequately controlled. For instance, core/shell,[3-8] yolk-shell[9-11] and heterodimers[12-16] are well-known complex systems that have shown enhanced optical, catalytic and magnetic properties compared to their individual single-component materials.

Amongst these nanomaterials, FePt-based nanocrystals attract special attention of researchers due to their potential application in storage medium or as permanent magnets. Moreover, it has been demonstrated that, due to their low toxicity[17] and high degree of functionalization,[13,18-19] the FePt nanoparticles (NPs) are also suitable for biomedical applications such as magnetic separation, hyperthermic ablation and $T_2$ MRI contrast agents.[20-22] It is well known that $Fe_3O_4$ is one of the most suitable for the later purposes. Thus, a multicomponent FePt-$Fe_3O_4$ NPs would be a promising material that could take advantages of the intrinsic properties of the both ends.

Recently, some efforts have been focused to synthesize FePt-based core-shell or heterodimers to obtain soft-hard magnetic nanostructures that display exchange-spring coupling effects.[16,23-27] However, in order to obtain the FePt $L1_0$ phase with hard magnetic properties, it is usually necessary to perform a thermal annealing of the NPs, even though this may cause some disadvantages like grain growth or coalescence. Several works have shown that the post-annealed product from FePt-based core-shell or heterodimers nanocrystals depends on factors like the Fe:Pt ratio, the type of surfactant or annealing atmosphere.[28-29] The knowledge of the kinetic process that takes place during annealing is relevant for fundamental studies but also for possible applications when a controllable magnetic response is required. Thus, a detailed investigation of the thermal evolution of FePt-based bi-component samples is still necessary to obtain valuable information on the role of intermediate phases and transformations involved during the annealing.

To this aim, here we investigate the thermal evolution of FePt-based metallic/magnetic nanocrystal prepared by one-pot one-step chemical synthesis method. We study their magnetic properties and investigate the stability of the as-made nanocrystal having a noncentric geometry. Our investigation is focused on the thermal evolution of heterodimers during the annealing under an inert atmosphere. The thermal evolution was followed by *in situ* time-resolved Pt $L_3$ and Fe K-edges x-ray absorption near edge spectroscopy (XANES) experiments. Throughout a semiquantitative analysis of XANES spectra we are able to follow the evolution of the phases involved during the annealing. The sample characterization is also complemented by using x-ray diffraction (XRD), transmission electron microscopy (TEM), extended x-ray absorption fine structure (EXAFS), Mössbauer spectroscopy and magnetometry. Our results demonstrate the important role played by the different reduction processes that take place during the heat treatment. This fact as well as the effect of the absence of a surfactant interface between heterodimers´ domains is discussed to explain the stabilization of a soft magnetic phase. We believe that this fundamental study give a pathway to obtain exchange-coupled materials with tunable properties, i. e., throughout the control of reducing mechanisms and interface influence.

## 2. Experimental methods

### 2.1 Synthesis

The NPs have been synthesized by a high-boiling coordinating solvent method.[13] The procedure involved heating 0.5 mmol of platinum(II) acetylacetonate (Pt(acac)$_2$) (Sigma-Aldrich, 99.99%) in 15 ml of 1-octadecen (Alfa Aesar, tech 90%) under N$_2$ flux up to 120 ºC, followed by the injection of 2.5 mmol of oleic acid (Sigma Aldrich, tech. 90%) and oleyamine (Sigma Aldrich, tech 70%) and 3 mmol of iron pentacarbonyl Fe(CO)$_5$ (Sigma Aldrich). The balloon flask was then sealed to avoid oxidation and kept under a blanket of nitrogen. The solution was slowly heated at a rate of 5 K/min up to 320 ºC and refluxed for 30 min. Afterwards, the black dispersion was cooled down to room temperature (RT). The obtained NPs were cleaned by cyclic precipitation and redispersed with ethanol (Panreac, absolute PRS) and hexane (Alfa Aesar, HPLC grad 95%). Finally, the particles were stored in 25 ml hexane with 0.05 ml oleic acid and oleylamine in a gloves chamber.

The concentration of iron and platinum atoms was determined with elemental analysis. For this purpose samples were digested with HNO$_3$ to oxidize the organic coating and then, with HCl to dissolve the metals. The Fe:Pt concentration was then measured in an induced coupled plasma emission spectrometer (ICP-ES) PERKIN ELMER OPTIMA 2100 DV. The result gives a Fe:Pt mean composition 67:33.

The as-synthesized NPs were thermally treated under inert atmosphere up to 873 K for different experiments (see bellow).

### 2.2 XRD and TEM characterization

Phases, crystal parameters and grain size were determined by RT XRD measurements, which were carried out with a Siemens diffractometer D5000, using Cu

K radiation. The particle sizes were characterized by TEM micrographs in a 200 keV JEOL-2000 FXII microscope. A drop of the dispersed suspension was placed onto a copper grid covered by a carbon film. The mean particle size, *d*, and the logarithmic standard deviation, σ, were obtained from digitized TEM images by counting more than 200 particles. Because the NPs are not spherical, the maximum Feret's diameter was used to compute the size, i.e., the maximum perpendicular distance between parallel lines which are tangent to the perimeter at opposite sides.

### 2.3 X-ray absorption experiments

The experiments were performed in the multibunch mode at D04B-XAFS-1 beamline of the Laboratório Nacional de Luz Síncrotron (LNLS), Campinas, Brazil. To take the absorption spectra a few drops of the dispersed suspension was placed onto a boron nitride pellet. All spectra were collected at RT in transmission mode using a channel cut monochromator (Si 111). Three gas-filled ionization chambers were used in series to measure the intensities of the incident beam, the transmitted beam through the sample and the beam transmitted by a reference foil in order to calibrate accurately the energy. The primary vertical slit size was about 0.3 mm, this yielded a resolution of about 2.0 eV on Fe K edge and close to 6.8 eV on Pt $L_3$ edge. The EXAFS signal $\chi(k)$ was extracted using the Athena and analyzed using the Arthemis software package. Simultaneous EXAFS fits of samples and references were performed (for details, see Supporting Information Section 1 (SI_1)). Pre-edge XANES fits at Fe-K edge were performed following the procedure described by Wilke *et al.* [30]

*In situ* XANES spectra at Fe K and Pt $L_3$ edges were collected in transmission mode using the D06A-DXAS beamline at LNLS.[31] The XANES experiment was performed in a controlled atmosphere cell heating the sample at a rate of 3 K/min in a flow of inert atmosphere (He) from ambient temperature up to 873 K and collecting synchronously mass spectrometry (GSD 301 Omnistar) data of the exhaust. The XANES data analysis was performed by subtracting a linear background and rescaling the absorbance by normalizing the difference between the baseline and the post-edge absorption in a region approximately 300 eV behind the edge. Factor Analysis (FA) was performed by using the FACTOR program.[32-35]

**2.4 Mössbauer and magnetic characterization**

The as-made and thermally treated samples were characterized magnetically by means of Quantum Design SQUID and Vibrating Sample Magnetometer. Hysteresis loops were taken at 10 and 300 K and at maximum applied field of 5 T. The thermal behavior was determined by zero-field-cooled (ZFC) and field-cooled (FC) measurements at 100 Oe applied field. For these measurements, the as-made NPs solution was dried under an inert argon atmosphere and weighted.

$^{57}$Fe Mössbauer spectra at 300 and 30 K were recorded in transmission geometry with a 512-channel constant acceleration spectrometer and a source of $^{57}$Co in Rh matrix of nominally 25 mCi. Isomer shifts (IS) are referred to metallic α-Fe at RT. The 30 K temperature was achieved using a Displex DE-202 Closed Cycle Cryogenic System. Mössbauer spectra were fitted using Recoil software package.[36]

**3. Results and Discussion**

**3.1 As-made NPs**

The XRD pattern (Fig. 1) shows the Bragg reflections belonging to an iron oxide with cubic structure in addition to those of an *fcc* metal compound (Pt) with a cell parameter $a$= 3.92 Å. The grain size calculated using the Scherrer´s equation is 5±1 nm for the metal compound and 7±1 nm for the oxide.

Figure 2(a) shows a representative TEM micrograph of the as-made NPs. The particles are heterodimers with a nonsymmetric snowmanlike shape, showing a large body and a small head, the latter exhibiting a comparatively darker image contrast. The dark region corresponds to a higher electron density likely associated to the metallic component, whereas the lighter one might be related to the presence of the oxide phase.[13,16] Figure 2 (b) shows the estimated mean particle size $d$ and standard deviation σ for both, the heterodimer head and body. By counting more than 200 NPs, the average values are $d$ = 4.8 nm and σ = 0.7 nm for the head, and $d$= 8.5 nm and σ=1.5 nm for the body. These values agree well with those determined by XRD. The smaller size obtained from XRD can be due to atoms at the particle surface that do not diffract coherently.

Figure 3 shows the Fe K-edge and Pt $L_3$ XANES spectra of as-made NPs. XANES features at the Fe K-edge mainly resembles those corresponding to magnetite $Fe_3O_4$. The pre-edge energy position is compatible with a $Fe^{3+}$-$Fe^{2+}$ mixture.[30] The main component of the pre-edge peaks of $Fe_3O_4$ arises from tetrahedrally coordinated $Fe^{3+}$ and the shoulder corresponds to $Fe^{2+}$-$Fe^{3+}$ ions octahedrally coordinated. The average oxidation state of iron from these XANES data is 2.5(1)+ , which is slightly less than expected for $Fe_3O_4$ (2.67+). This value indicates that the major iron phase is magnetite but the sample might contain a small percentage of a metallic phase (see below). From this result the amount of the other spinel iron phase maghemite γ-$Fe_2O_3$ regularly found in iron oxide NPs, if any, is quite low.[37]

Fe K-edge EXAFS are acceptably fitted to a cubic spinel structure (see Table 1). The Fe-O distances are close to the values of 1.89 and 2.06 Å expected for the tetrahedral and octahedral sites in $Fe_3O_4$, respectively. A reduction of the amplitude of the second main peak is observed when compared with the Fourier transform of bulk $Fe_3O_4$, which is probably related to the local structural disorder due to the higher percentage of atoms at the particle surface layer in as-made NPs.

We observe that Pt $L_3$-edge XANES of the as-made NPs and Pt reference foil are similar, suggesting that the Pt local structure mainly resembles that corresponding to Pt metal rather than to a FePt alloy.[38] A more detailed analysis can be extracted from the Fourier transforms $\chi(R)$ of the EXAFS spectra at the Pt $L_3$ edge shown in Fig. 4 (fitting results are shown in Table 1). The nearest neighbor distance Pt–Pt (or Pt-Fe) is close to 2.74 Å, which is shorter than the value expected for pure platinum (2.77 Å). This would indicate the incorporation of low quantity (*circa* 10 at.%) of iron into platinum forming a FePt alloy. This is also supported by the presence of a small phase shift on $\chi(k)$ signals (see SI_2). Hereafter we will refer to this component as a Pt-rich FePt head of the heterodimer. Reports on equiatomic FePt NPs with a Pt-rich core and Pt-depleted shell can often be found in the literature.[38-41] The explanation on the inhomogeneity in the Pt concentration lies in the different nucleation velocity of both atomic species: Pt atoms nucleate faster than the Fe ones, leading to a Pt concentration that decreases from inside to outside of the NP.

The Mössbauer spectrum of the as-made NPs showed an absorption signal only below the fusion temperature of oleic acid $T_f$ = 281 K. Spectrum taken at 250 K (not shown) mainly consisted of a broad doublet signal. At 30 K the spectrum (Fig. 5) shows a resolved magnetic signal (relative area percentage about 90%) plus a non-magnetic or relaxing signal with $IS$ = 0.37 mm/s that represents unblocked iron oxide particles. A small doublet with a relative area approximately 5% was also included ($IS$= 0.03 mm/s), which can be assigned to a metallic minority phase. The magnetic signal needs to be fitted with at least two sextets but the lack of resolution due to the broad lines and the poor signal/noise ratio makes the determination of proportions and parameters of each component difficult. The best fit was obtained assuming one sextet with hyperfine field $B_{hf}$ = 519 kOe and $IS$ = 0.48 mms, and another one with $B_{hf}$ = 491 kOe and $IS$= 0.42 mm/s. Both sextets are assigned to blocked $Fe_3O_4$ NPs, but γ-$Fe_2O_3$ cannot be ruled out.[42-43] Further, these results indicate that, within the Mössbauer window time, the heterodimer´s body composed by iron oxide behaves superparamagnetically at ambient temperature.

A quite common difficulty found in several investigations about nanosized iron oxides is related with the distinction between $Fe_3O_4$ and γ-$Fe_2O_3$ either whether these oxides are the only components[37,44] or when they are chemically bounded to other materials.[45] Indeed, both oxides have the same spinel structure and similar saturation magnetization, being the latter affected by size and/or surface effects.[46] Also, the Mössbauer hyperfine parameters are modified with respect to those belonging to their bulk counterparts and the contribution of each oxide can only be estimated by taking spectra under an applied magnetic field.[47] In our case, even though the Mössbauer and XRD results are not conclusive, the oxidation state determined by XANES and EXAFS results suggest that the oxide part consists of $Fe_3O_4$.

The magnetic saturation $M_S$ at 10 K is 30 emu/$g_{NP}$ (per gram of NP), whereas at ambient temperature the saturation is about 20 emu/$g_{NP}$ (see Fig. 6). Considering the average composition of as-made heterodimers ($Fe_{67}Pt_{33}$), $M_S$ per gram of $Fe_3O_4$ can be roughly estimated as 68 emu/$g_{oxide}$. This $M_S$ value is lower than the $M_S$ of bulk $Fe_3O_4$ (90 emu/g) and can be ascribed to the spin disorder at the surface of the NPs. The room temperature $M$-$H$ loop shows small hysteresis (coercive field $H_C$ = 10 Oe), increasing at 10 K to $H_C$ = 150 Oe. The departure of the FC magnetization curve from the ZFC one, characteristic of the presence of moments getting blocked, occurs at about 130 K (Fig. 6). The blocking temperature $T_B$ corresponding to the maximum of the ZFC curve is about 110 K. The broad ZFC maximum (Fig. 6) suggests that there are strong magnetic dipole-dipole and/or exchange interactions among the particles. Thus, our results indicate that the as-made NPs consist of a heterogeneous distribution of atoms forming a metallic-oxide Pt-rich FePt/ $Fe_3O_4$ heterodimer.

It is important to highlight that our synthesis procedure consists in adding the platinum salt and the iron pentacarbonyl to the solution in one step at the beginning of the synthesis. This route slightly differs from others previously reported,[13,16,23] in which first FePt seeds or NPs are formed and afterwards more iron pentacarbonyl or iron oxide NPs are added to get bimagnetic FePt-iron oxide NPs. We believe that in our case, due to the fact that Pt atoms nucleate faster than the Fe ones, the nucleation of FePt NPs takes place during the initial stages of the reaction. Once the Pt is exhausted, the remaining iron atoms grow on a surface defect of FePt giving place to the iron oxide body of the heterodimer. The fact that the oxidized body consist of $Fe_3O_4$ would indicate that the thermal decomposition of organic precursors in 1-octadecene solvent operate as stabilizer of magnetite.[46] In addition, the storage in an inert atmosphere to avoid exposing to air has prevented the heterodimers from oxidation. Recently, Corrias *et al.*[37] demonstrated that the smaller the sizes of iron oxide NPs the component $Fe_3O_4$ becomes more stable, i. e., the risk of oxidation decreases.

Summarizing, our NPs can be visualized as two different entities attached to each other but without border (like Siamese). Apart from differences in size and shape, the morphology of these NPs differs from that previously reported,[16,23] where NPs are composed of two different entities, probably attached by the surfactant, but with a well defined frontier or boundary.

### 3.2 Thermal evolution of Pt-rich FePt/Fe$_3$O$_4$ heterodimers

To follow the thermal evolution of these Pt-rich FePt/Fe$_3$O$_4$ heterodimers we have performed *in situ* XANES experiments at Pt L$_3$ and Fe K-edges. We observed that Fe-K absorption edge position (see Fig. 7) shifts monotonically towards lower energy values along the treatment, indicating a progressive reduction of Fe-containing phases. The Fe K-edge intensity of the white line (W-L) centralized at about 7130 eV progressively decreases as the annealing temperature increases, showing a relative maximum near 440 K (see inset Fig. 7). In addition, an abrupt intensity decrease of W-L occurs at about 840 K. These changes are concomitant with the maximum values of H$_2$ and CO detected by mass spectrometry (See SI_4). A similar behavior at the same temperatures was observed for the intensity of W-L line corresponding to the Pt L$_3$-edge that corresponds to the electron transition from core 2*p* electrons to unoccupied 5*d* states (not shown). A decrease of the W-L could be either due to lower Pt 5*d* vacancies or to size effects.[48-49]

To further characterize the thermal evolution of FePt/Fe$_3$O$_4$ heterodimers, a set of Fe K-edge XANES spectra has been subjected to Factor Analysis (FA). The hypothesis behind FA is that a large data set (*D*) consists in a linear combination of a limited number of *n* factors. Thus, the large data set containing *r* spectra each having *c* data points can be reproduced by two matrices, one of them (*S*) containing the basic spectral factors and the other one (*C*) containing the corresponding concentration profile of the factors in each of the original spectra, with *E* representing the residual or error matrix,

$$D_{rxc} = C_{rxn} \cdot S_{nxc} + E_{rxc}$$

The goal is the determination of the number *n* of factors (or components) in the data set which corresponds to the rank of the data matrix used in the analysis. Therefore, evolving factor analysis (EFA)[50] was used to determine that the number of factors can be fixed to three. Details of this analysis can be found in the SI_3.

Figure 8 shows the abstract row matrix (spectral factors), with the first five columns represented, giving visual evidence that the abstract factors number one, two, and three are the main signals. Fourth and fifth abstract factors are dominated by low frequency background errors. Once the number of factors is established, the abstract $C$ and $S$ matrices can be obtained. Their adequate rotation by the iterative transformation factor analysis (ITTFA) gives the final solution, which is represented in Fig. 9(a) ($C_{rxn}$ matrix) and Fig. 10 ($S_{nxc}$ matrix).

Figure 9(a) shows the three spectral components determined by FA, while Fig. 9(b) shows XANES taken under the same experimental conditions to tentatively identify these components. Spectra 2 and 3 can be clearly linked to initial and final iron species - i.e., as-made Pt-rich FePt/$Fe_3O_4$ heterodimers and $Fe_3Pt$ (see discussion below), respectively- assumed as the single phases at these stages. Even though the assignment of spectrum 1 is not straightforward, a simple inspection allowed us to qualitatively establish that their spectral factor has similar features with XANES of iron oxide FeO (wüstite) (Fig. 9). Figure 10 shows the result of FA by considering some selected spectra. At the first stages, the percentage of $Fe_3O_4$ decreases at expenses of component 1. Above 550 K, the $Fe_3Pt$ phase starts to be formed with the consequent detriment of the iron oxides phases. We observe that the percentage of component 1 displays an abrupt peak at about 843 K, coincidently with a marked W-L detriment (see Fig. 7). This temperature is around the eutectic point where FeO-$Fe_3O_4$ and Fe coexist and a temperature increase strongly favors the reduction of FeO to a metallic phase.[51] This behavior reinforces our assumption that this component 1 would represent the formation of wüstite as an intermediate phase in the reduction process that takes place during the thermal annealing.

It is well known that fcc FePt NPs transform to the hard magnetic phase above 820 K.[52] Therefore, the fact that $Fe_3Pt$ begins to nucleate at 550 K suggests that the energy of formation of $Fe_3Pt$ from Pt-rich FePt/$Fe_3O_4$ heterodimers is lower than the energy needed to transform FePt NPs from cubic to tetragonal symmetry.

### 3.3 Heterodimers post-annealing: $Fe_3Pt$ stabilization

XRD reflections of the annealed heterodimers can be indexed to an $Fe_3Pt$ intermetallic compound with a $Cu_3Au$-type structure ($L1_2$) with a cell parameter $a$= 3.74 Å.[53] The narrowing of XRD peaks with respect to the as-made sample shows the improvement of the crystallinity. The average grain size estimated by using the Scherrer´s equation is 80 nm.

The Fe K-edge XANES spectrum showed a metallic pre-edge feature (Fig. 3). Both, the W-L and post-edge region features resemble those of ordered $Fe_3Pt$.[54] Simultaneous fits results of Fourier transform of EXAFS signals at Pt $L_3$ and Fe K-edges without phase correction (Fig. 11, Table 2) indicate that the Pt and Fe environments coincide with those expected for Pt and Fe in the $Fe_3Pt$ compound.[55]

The RT Mössbauer spectrum can be decomposed into a central singlet (*IS*= 0.08 mm/s, relative area 0.9) and a weak $Fe^{2+}$ signal (*IS*= 0.87 mm/s), the latter probably related to the residual FeO also indicated by FA. The spectrum taken at 30 K mainly shows a broad magnetic contribution with $B_{hf}$ = 362 kOe and *IS*= 0.33 mm/s, which we assign to iron at $Fe_3Pt$ sites.[56] The difference between FC and ZFC magnetization curves starts above RT and increases monotonically when lowering the temperature (see inset Fig. 12). The ZFC maximum occurs at a temperature above ambient. The *M-H* curve measured at 298 K presents hysteresis with $H_C$ = 75 Oe, while at T= 10 K, $H_C$ = 185 Oe. $M_S$ at 10 K was found to be about 120 emu/$g_{NP}$, being this value close to the magnetic moment reported for $Fe_3Pt$.[53]

The present results show that the reduction of the iron oxide phase plays an important role that favours the interdiffusion of Pt and Fe atoms. This statement might be explained as follow. Heat treatment in inert atmosphere leads to the surfactant decomposition that releases C, CO and $H_2$ producing a stark reducing atmosphere (see Fig. 6 in SI_4). A first process, that takes place from RT to about 550 K, is the progressive reduction of $Fe_3O_4$ that transforms to FeO (see Fig. 10). Above this temperature, the release of more hydrogen and hydrocarbons further reduces $Fe^{2+}$ and $Fe^{3+}$ to $Fe^0$ and give rise to the $Fe_3Pt$ phase. The latter implies that Fe and Pt interdiffusion has started to take place. Up to about 725 K, the amount of $Fe_3Pt$ increases at expense of further reduction of magnetite and wüstite and the resulting interdifussion. By increasing the annealing temperature, $Fe_3O_4$ continuously transforms to FeO whereas the amount of the metallic component remains almost constant. The $Fe_3O_4$ to FeO transformation reaches a maximum near 840 K, a temperature close to the eutectic point where FeO-$Fe_3O_4$ and Fe coexist. Afterwards, the remaining $Fe^{2+}$ reduces to $Fe^0$ and the interdiffusion is once more activated. Thus, as the annealing temperature reaches 840 K, the nucleated intermetallic phase further drives the complete transformation and frustrates the formation of a hard-soft nanocomposite.

The interdiffusion process is probably helped by the fact that the coexisting domains are continuously attached. Another mechanism would take place when FePt and Fe oxide NPs are bounded only by surfactants, .i.e, there exists physical frontiers between the particles.[16,23,57] In these cases, the Pt atoms would need to overcome the energy barrier of the interface FePt/Fe oxide to form $Fe_3Pt$. As the thermal energy is lower than the interface energy barrier, the Pt atoms would be limited to spread out in the same FePt NP, favoring the phase transformation of FePt from cubic to tetragonal structure. Once the thermal energy matches the interface energy barrier, the Pt atoms are free to move to the iron oxide NPs to finally form $Fe_3Pt$. This assumption about the function of the frontiers of the heterodimers ends needs to be confirmed by performing new experiments.

## 4. Summary and Conclusions

Surfactant coated iron platinum NPs with molar ratio 67:33 were synthesized by a wet chemical one-pot one-step method, giving rise to a heterogeneous distribution of atoms that forms a metallic-oxide Pt-rich $FePt/Fe_3O_4$ heterodimer, which behaves superparamagnetically at RT. Due to the synthesis procedure, both ends of the heterodimer are attached to each other without borders.

As-made NPs were subjected to thermal annealing under an inert gas flow and their evolution was followed by means of *in situ* XANES experiments. Results from FA analysis of these spectra showed the progressive reduction of the sample. During the first stages, a progressive reduction of the type $Fe_3O_4 \rightarrow 3\ FeO + O$ of the heterodimers' body takes place. The metallic part showed an increase of the *d*-electron density of Pt atoms. The presence of reducing agents, such as C, CO, and $H_2$ released from the thermolysis of oleic acid and oleylamine might contribute to produce changes in the oxidation state of the heterodimers. At about 440 K the iron oxide reduction produces a relatively high percentage of FeO. Once FeO becomes the majority oxide phase and at temperatures above 550 K, the interdiffusion between Pt and Fe atoms from Pt-rich FePt and $Fe_3O_4$/FeO components, respectively, begins and nucleates the $Fe_3Pt$ compound. The amount of nucleated $Fe_3Pt$ phase remains almost unchanged up to about 840 K -close to the temperature at which commonly the FePt soft to hard transformation occurs- where an abrupt increase of FeO percentage is observed. This second reduction step further drives the almost complete phase transformation to stabilize the $Fe_3Pt$ phase. Magnetically, the post-annealed sample shows soft properties with a high saturation magnetization. These results show that the reduction of magnetite and the emergence of wüstite as intermediate phase act as a catalyst that facilitates Fe and Pt interdiffusion to form $Fe_3Pt$ instead of exchange coupled $FePt/Fe_3Pt$.[57] The interdiffusion is also helped by the fact that the metallic and oxide domains are continuously joined.

We conclude that inhibition of the Fe and Pt interdiffusion throughout the control of reducing mechanisms and interface influence should be of critical importance to synthesize tunable exchange-coupled materials for specific purposes. On the other hand, applications in the biomedicine field (drug delivery, hyperthermic ablation or MRI) would require non-corrosive superparamagnetic materials with a large magnetization at RT. Under these requirements, annealed Pt-rich $FePt/Fe_3O_4$ heterodimer could be a promising candidate for these applications due to its larger magnetization in comparison with iron oxides, low corrosivity and large stability, provided the coalescence and grain growth can be adequately controlled.

**Acknowledgements**

We appreciate financial support by LNLS, Campinas, SP, Brazil (proposals D06A - DXAS – 7739 and D06A–DAXS–6669) and CONICET, Argentina. This work was supported by grants from the Spanish Ministry of Science and Innovation (CSD2007-00010, MAT2009-14741-C02-00) and the Madrid regional government CM (S009/MAT-1726). SJAF thanks M. Fernández-García and S. Pascarelli for fruitful comments on the manuscript. We thank F. Zambello (LNLS) and E. J. de Carvalho (LNLS) for their for his kind assistance during the DXAS experiments and G. Azevedo, C. Rodríguez Torres and M. Ceolín for the beam time provided to perform the EXAFS experiments. PP thanks to MP Morales for TEM images and ICP characterizations.

**Supporting Information Available:** Additional data about the EXAFS data analysis, experimental EXAFS characterization at Pt-$L_3$ edge of as-made sample, the Factor Analysis performed on Fe K-edge XANES and the mass spectroscopy results are available. This information is available free of charge via the Internet at http://pubs.acs.org.

**Table 1**: EXAFS Fe $K$-edge and Pt $L_3$-edge fitted parameters of Pt-rich FePt/Fe$_3$O$_4$ heterodimers. $N$ is the coordination number, $R$ is the average distance to the central atom, $\sigma^2$ is the Debye-Waller factor and $E_0$ is the energy shift.

| Sample | Edge | Shell[a] | N | R (Å) | $\sigma^2$ (Å$^2$) x $10^{-2}$ | $E_0$ (eV) |
|---|---|---|---|---|---|---|
| As-made NPs | $K$-Fe | O $_{Fe_3O_4}$ | 4.1 ±0.2 | 1.890 ± 0.007 | 4.3 ± 0.7 | -7.2 ± 1.3 |
| | | O $_{Fe_3O_4}$ | 6.3 ± 0.2 | 2.020 ± 0.006 | 5.2 ± 0.7 | -7.2 ± 1.3 |
| | $L_3$-Pt | Pt$_{(1-x) Pt}$[b] | 9.4 ± 1.2 | 2.744 ± 0.005 | 0.64±0.08 | 6.6 ± 1.2 |
| | | Fe$_{x Pt}$[b] | 9.4 ± 1.2 | 2.744 ± 0.005 | 0.64±0.08 | 6.6 ± 1.2 |

[a] Model compound used to obtain the phases. [b] The value obtained for the iron dopant is x=0.1±0.05.

**Table 2**: EXAFS Fe-K and Pt L$_3$-edge fitted parameters of Pt-rich FePt/Fe$_3$O$_4$ heterodimers (post-annealed NPs). $N$ is the coordination number, $R$ is the average distance to the central atom, s$_2$ is the Debye-Waller factor and E$_0$ is the energy shift.

| Sample | Edge | Shell[a] | N | R (Å) | $\sigma^2$ (Å$^2$) x $10^{-2}$ | $E_0$ (eV) |
|---|---|---|---|---|---|---|
| Post-annelled NPs | $K$-Fe | O$_{FeO}$ | 5.4 ± 1.5 | 1.97 ± 0.11 | 4.5 ± 3.2 | 6.7 ± 4.2 |
| | | Pt$_{Fe_3Pt}$[b] | 3.6 ± 1.5 | 2.59 ± 0.01 | 0.9 ± 0.1 | 6.7 ± 4.2 |
| | | Fe$_{Fe_3Pt}$[b] | 7.2 ± 1.5 | 2.59 ± 0.01 | 1.2 ± 0.1 | 6.7 ± 4.2 |
| | $L_3$-Pt | Pt$_{Fe_3Pt}$[b] | 10.8 ± 1.5 | 2.59 ± 0.01 | 11 ± 0.8 | 4.6 ± 4.7 |

[a] Model compound used to obtain the phases. [b] A third cumulant was used to fit all these shells. Values obtained were -4±2×10$^{-4}$.

**Figure Captions**

**Figure 1**: X-ray diffraction pattern of as-made heterodimers (a) and after annealing at 873 K in inert atmosphere (b).

**Figure 2**: a) TEM micrograph of as-made heterodimers; the scale bar is 10 nm. b) Size distribution of the heterodimer head and body.

**Figure 3**: XANES spectra at the Pt $L_3$ and Fe K-edges of as-made Pt-rich FePt/Fe$_3$O NPs. Spectra of Fe and Pt reference foils are also shown for comparison.

**Figure 4**: Fourier transforms of $k^3\chi(k)$ EXAFS at Pt $L_3$-edge and at Fe K edge of the as-made Pt-rich FePt/Fe$_3$O NPs. Solid lines correspond to the fitting results with parameters shown in Table 1. XANES of bulk Fe$_3$O$_4$ is also shown for comparison.

**Figure 5**: Mössbauer spectra at 30 K of the as-made Pt-rich FePt/Fe$_3$O NPs.

**Figure 6**: (a) M vs H at 5 and 300 K, the inset shows a detail of the hysteresis curve at 10 K. (b) M vs T curves as-made heterodimers NPs, the applied field is 100 Oe.

**Figure 7**: XANES spectra at Fe K-edge in thermal evolution treatment in He atmosphere of Pt-rich FePt/Fe$_3$O$_4$ heterodimers. The inset shows the white line evolution along the treatment.

**Figure 8**: Abstract spectral factors.

**Figure 9**: (a) Predicted pure components for FePt sample at Fe-K edge; (b) experimental XANES of iron phases to be compared with the predicted components along the treatment (component 1: FeO, component 2: as made NPs, component 3: post-annealed sample).

**Figure 10**: Thermal dependence of the percentage of iron species $Fe_3O_4$, FeO, and $Fe_3Pt$ obtained by performing a factor analysis (FA) of the Fe K-edge XANES data.

**Figure 11**: Fourier transforms of $k^3\chi(k)$ EXAFS data after treatment at Pt $L_3$-edge and Fe K edge.

**Figure 12**: M vs H of post-annealed NPs measured at 10 K and 300 K.

**Figure 13**: Mössbauer spectra of post-annealed NPs recorded at 293 and 30 K. Solid lines correspond to fitting results.

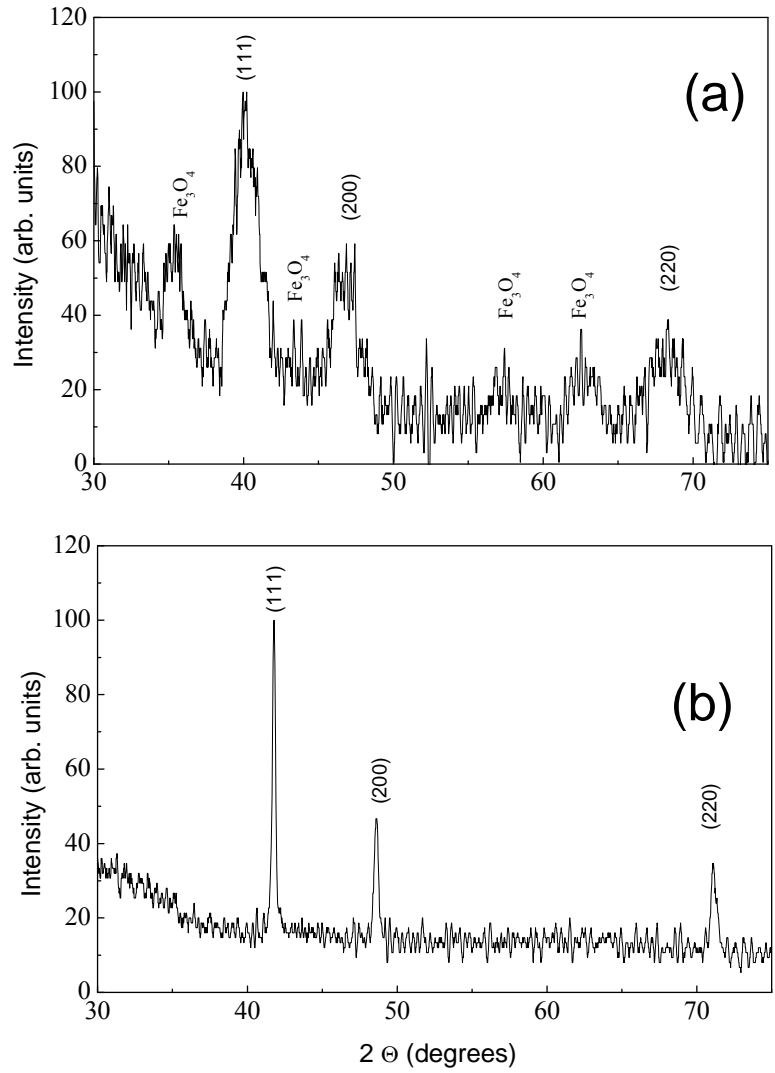

Figure 1

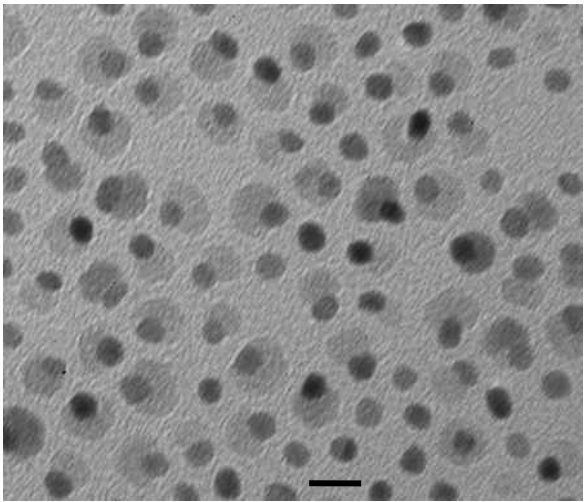 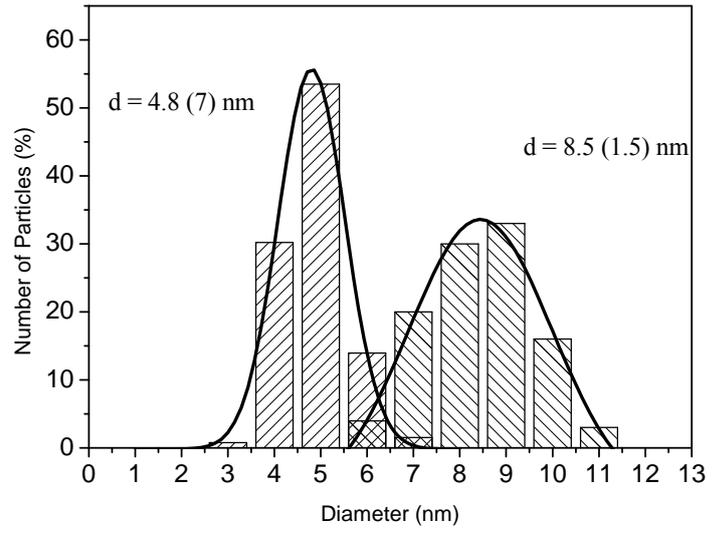

(a)             (b)

Figure 2

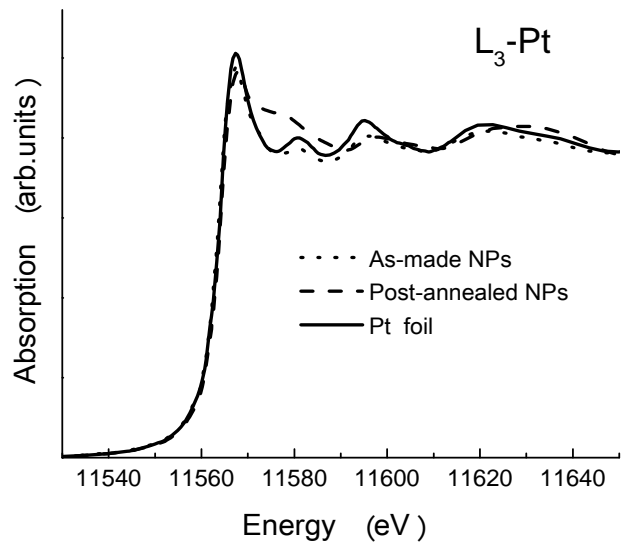

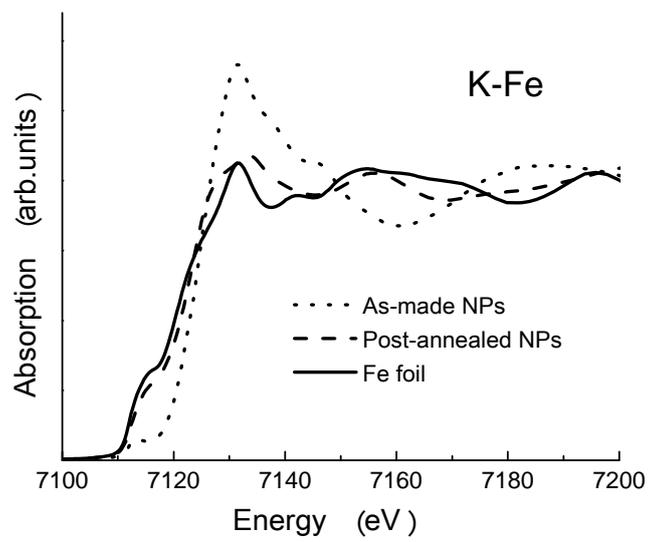

Figure 3

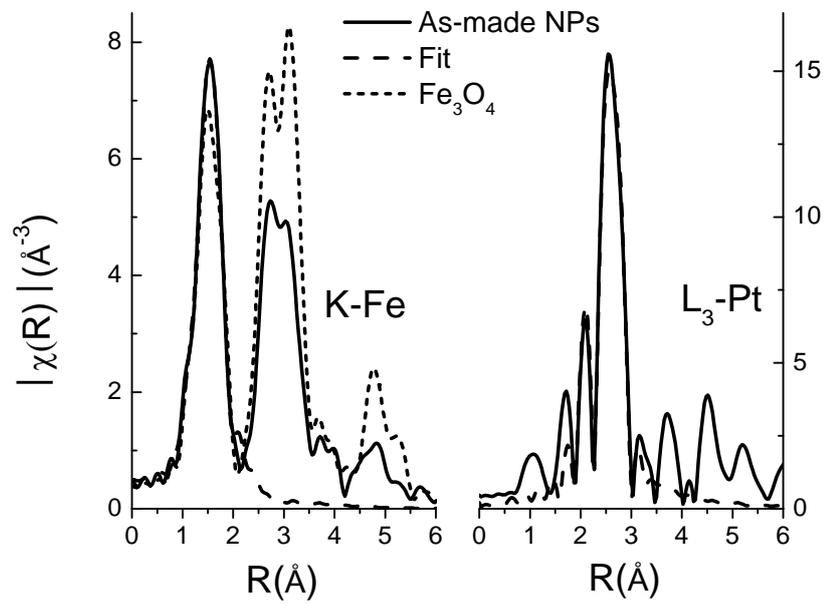

Figure 4

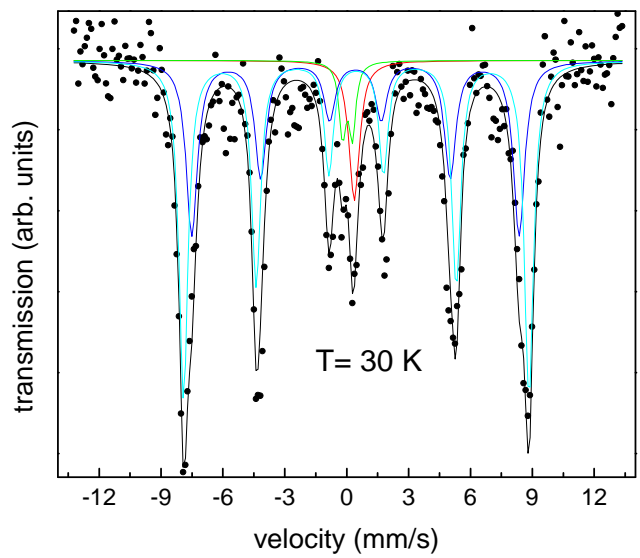

Figure 5

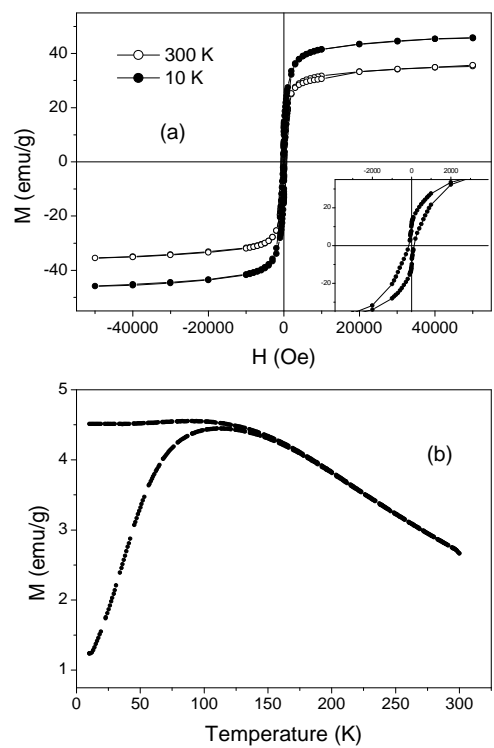

Figure 6

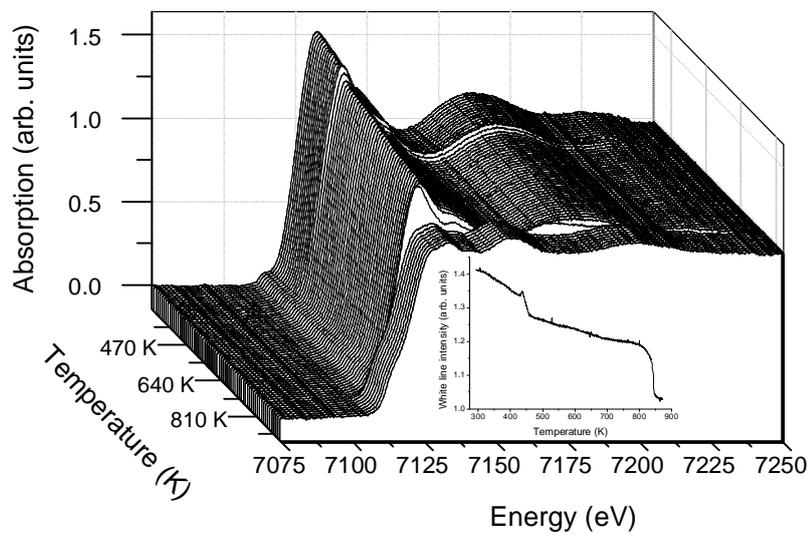

Figure 7

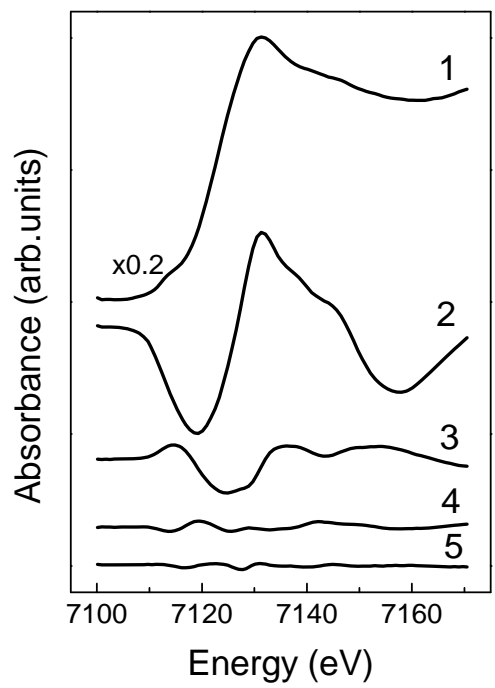

Figure 8

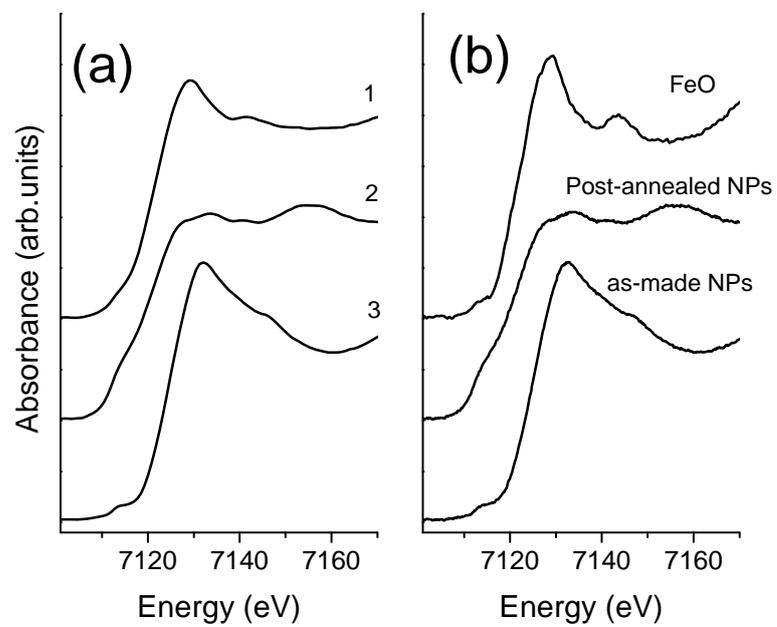

Figure 9

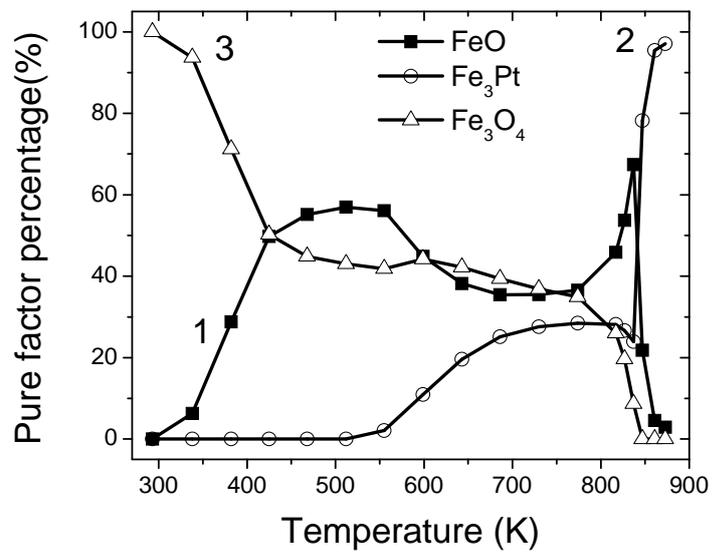

Figure 10

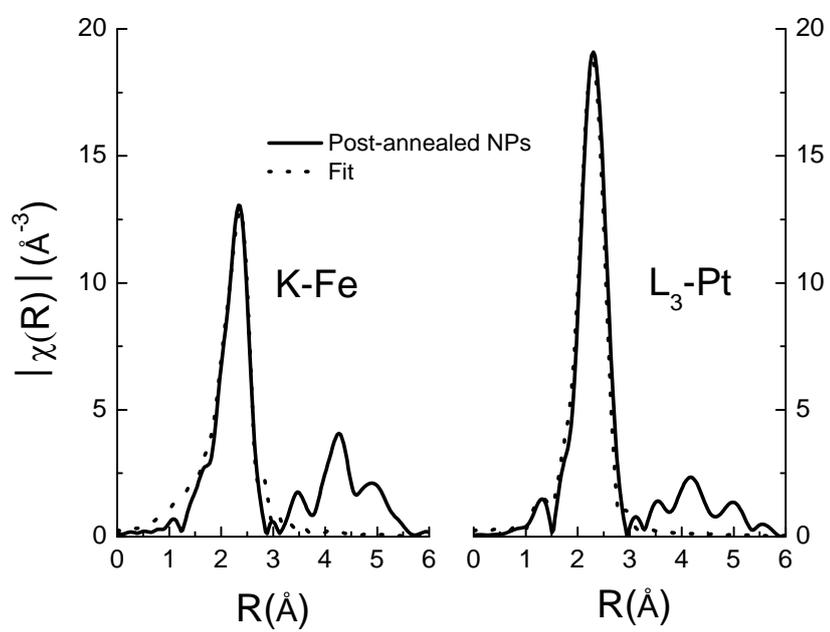

Figure 11

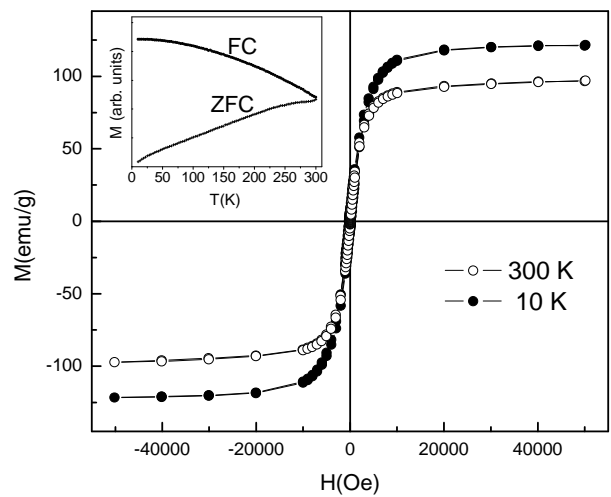

Figure 12

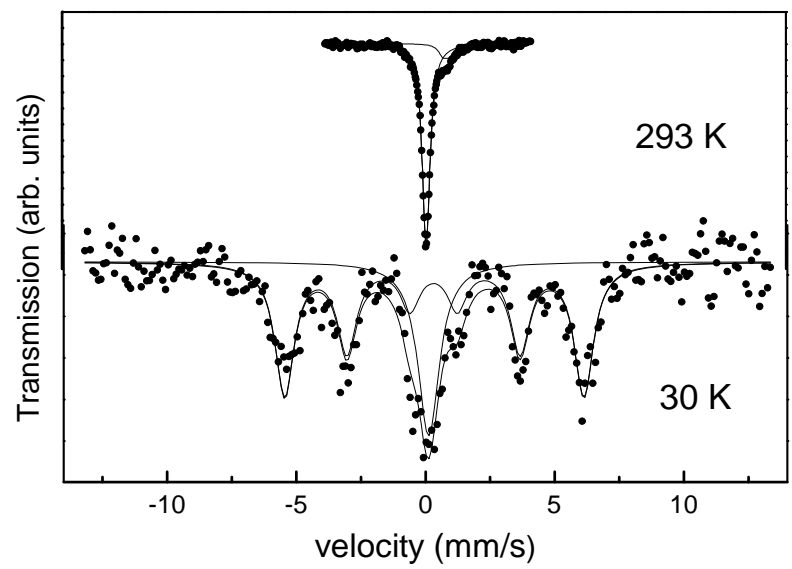

Figure 13

TOC

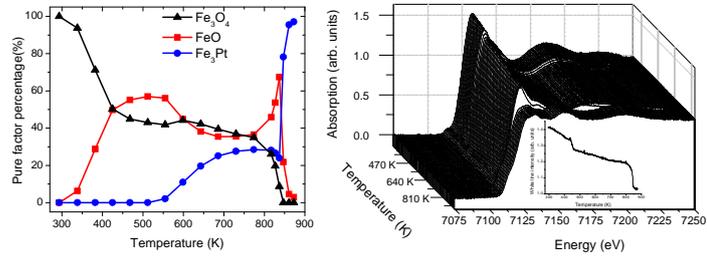